# Second Harmonic Generation in Deeply Sub-Wavelength Waveguides

V. Roppo[1,2,*], M. A. Vincenti[1,3], D. de Ceglia[1,3], M. Scalora[1]

[1]C. M. Bowden Research Center, AMSRD-AMR-WS-ST, RDECOM, Redstone Arsenal, AL 35898-5000, USA

[2]Laser Physics Centre, RSPhysSE, Australia National University, Canberra ACT 0200, Australia

[3]Aegis Technologies Inc., 410 Jan Davis Dr., Huntsville, AL 35806, USA

*Corresponding author: vito.roppo@gmail.com

## Abstract

We theoretically investigate second harmonic generation in extremely narrow, sub-wavelength semiconductor and dielectric waveguides. We discuss a novel guiding mechanism characterized by the inhibition of diffraction and the suppression of cut-off limits in the context of a light trapping phenomenon that sets in under conditions of general phase and group velocity mismatch between the fundamental and the generated harmonic.

Early studies of Second Harmonic (SH) generation made it clear that the emitted, nonlinear signal is composed of two different components [1]. This is a direct consequence of the adoption of the oscillator model for the bound electrons. The nonlinear Maxwell's equations are a set of inhomogeneous partial different equations and the general solution is then composed by the linear superposition of the homogeneous (HOM) solution and a particular solution obtained by including the driving, Fundamental Field (FF) term, i.e. the inhomogeneous (INH) solution [1-4]. Ever since Franken's experiment [5], only the HOM SH has been under the spotlight. Lately, however, a number of theoretical and experimental works have shed new light on the characteristics that make the INH SH component quite peculiar and useful in forbidden wavelength ranges [6, 7]. Briefly, the INH SH propagates with the complex index of refraction of the FF, regardless of material dispersion present at the SH (or third harmonic, for that matter) wavelength, provided the nonlinear coefficient do not vanishes. The harmonic field travels with the same group velocity as the fundament pulse [2, 3], and it is not absorbed even if it is tuned in a spectral region that is either highly absorbing [6] or metallic in nature [7]. Other experimental and theoretical details may be found in references [8, 9], where the behavior is demonstrated in two dimensional environments.

In practice these discoveries open new vistas on the subject of nonlinear frequency conversion: one should always be aware of the fact that when a FF impinges on a nonlinear material (assumed to be relatively transparent at the fundamental wavelength) it generates a HOM SH component that propagates according to material dispersion, and an INH SH component that mimics the behavior of the FF, for *any* phase and group velocity mismatches. These two parameters, however, dictate the relative and absolute intensities of the two SH components. In general, for large phase mismatch both components have similar amplitudes and are characterized by low conversion efficiency [6, 7, 8]. For example the INH SH component is generally resonant and enhanced in cavities designed to be transparent and resonant *only* at the FF [10].

In this letter we show that it is possible to design deeply sub-wavelength waveguides and generate INH SH signals circumventing traditional diffraction and cut-off limits. We consider a generic, infinitely long, linear background material where a nonlinear channel has been imprinted, as shown in Fig.1(a). We will refer to this region as a nonlinear waveguide. For simplicity, both the linear and nonlinear materials can be assumed to be GaP, whose dispersion is shown in Fig.1(b) [11, 12], where we can identify three regions: For $\lambda > 480$nm the material is *transparent*; for 260nm $\leq \lambda \leq$ 480nm the material is quite *absorptive;* for $\lambda <$ 260nm GaP is absorptive but the real part of the permittivity has negative values (*metallic region*). Let us now consider a FF pulse propagating along the waveguide, with beam waist much larger than the transverse dimension of the channel. This structure does not present any particular *linear* features, neither at FF nor at SH wavelengths, due to the absence of any index variation. However, what happens to the *generated* SH signal? The answer to this question is not unique and it depends on the relative tuning conditions of the FF and SH. In order to answer this question we perform a numerical analysis using a Fast Fourier Transform, time-domain beam propagation method [6, 9, 13].

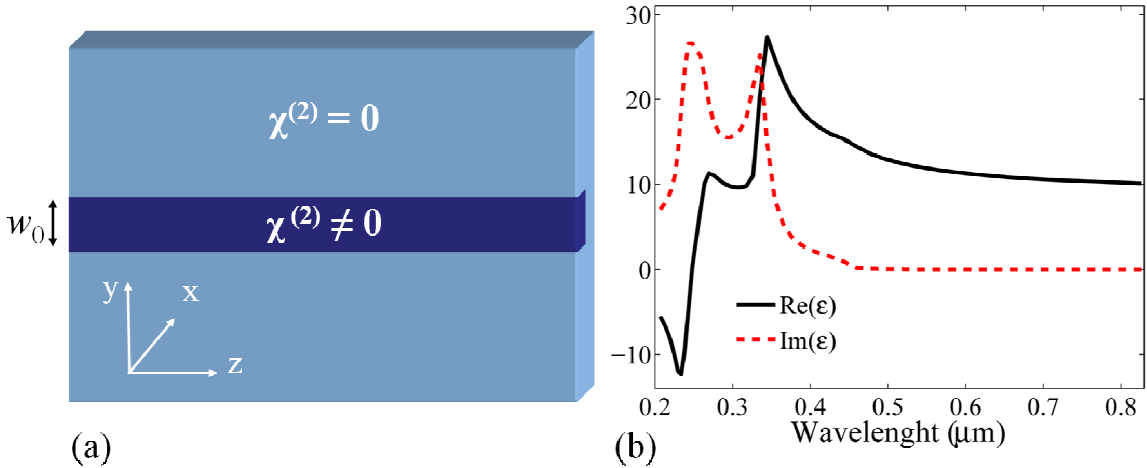

**Figure 1**. (Color online) **(a)** Sketch of the waveguide and excitation scheme. A nonlinear channel is created in a linear background material. The linear properties of the entire sample are assumed to be those of GaP. **(b)** Dispersion relation of GaP from [12].

We integrate a Maxwell-Lorentz system of equations in two dimensions, without resorting to either slowly varying or undepleted pump approximations. The numerical model is similar to that used in [6], with the addition of a transverse coordinate to include diffraction and propagation on a two-dimensional plane. We then assume a transverse magnetic (TM) polarized is incident from the left and generates similarly polarized harmonic fields (namely, E fields along x-axis). However, similar results are obtained for transverse electric (TE) polarized fields.

We consider a first scenario with one nonlinear channel $w_0$=1μm wide and refractive index identical in all regions. In Fig.2 (Media 1) we plot a snapshot of the FF propagation (top) and generated SH (bottom). The FF is tuned at $\lambda_{FF}$=1.3μm, so that both FF and SH are tuned in the transparency region. The FF has a Gaussian shape in both dimensions, it is 30fs in duration and several times wider than the waveguide. The intensity is of the order of 0.1GW/cm$^2$. In Fig.2 the pump pulse has already propagated 10μm inside the material so that the only noticeable effect is a pulse compression in the propagation direction due to the high index of refraction of the material ($n_\omega$=3 and $n_{2\omega}$=3.4). The generated SH behaves more peculiarly. The generated HOM SH component (spot marked with 1 in Fig.2) walks off and lags behind the pump pulse. Material dispersion causes the group velocity of the HOM signal to be smaller compared to that of the fundamental pulse. We estimate $v_{g\omega}$=c/3.15 and $v_{g2\omega}$=c/3.81 (calculated as $v_g \sim c/(n+\lambda\partial n/\partial\lambda)$ ). The spatial separation of the two pulses is evident because we are well inside the medium (pulses have propagated ~10μm from the surface). The HOM SH signal propagates unbounded and quickly diffracts away from the nonlinear channel because of its subwavelength size. The k-vector of this component has a modulus given by $k_{2\omega}n_{2\omega}$ (marked with 1 in Fig.3 (a)).

In Fig.2 we also identify the generated, INH SH signal (spot number 2). Unlike the HOM signal, the INH component propagates locked to the pump at the FF group velocity, while the modulus of its k-vector is given by $2k_\omega n_\omega$ (Fig.3(a), marked with 2) [4]. However, the signal inevitably experiences strong diffraction because the transverse dimension of the waveguide is of the order of the wavelength, and portions of its energy spill out into the linear region. We note that diffraction is limited only in the region where the $\chi^{(2)}$ is non-zero (assumed of the order of 10pm/V), as the INH SH locks to the FF. In contrast, those portions of INH SH that exit into the linear region become free to propagate and to recover the k-vector dictated by material dispersion, namely $k_{2\omega}n_{2\omega}$ (Fig.3(a), marked with 3). The SH signal then slows down and takes on a peculiar arrow-shaped trail (spot 3 in Fig.2).

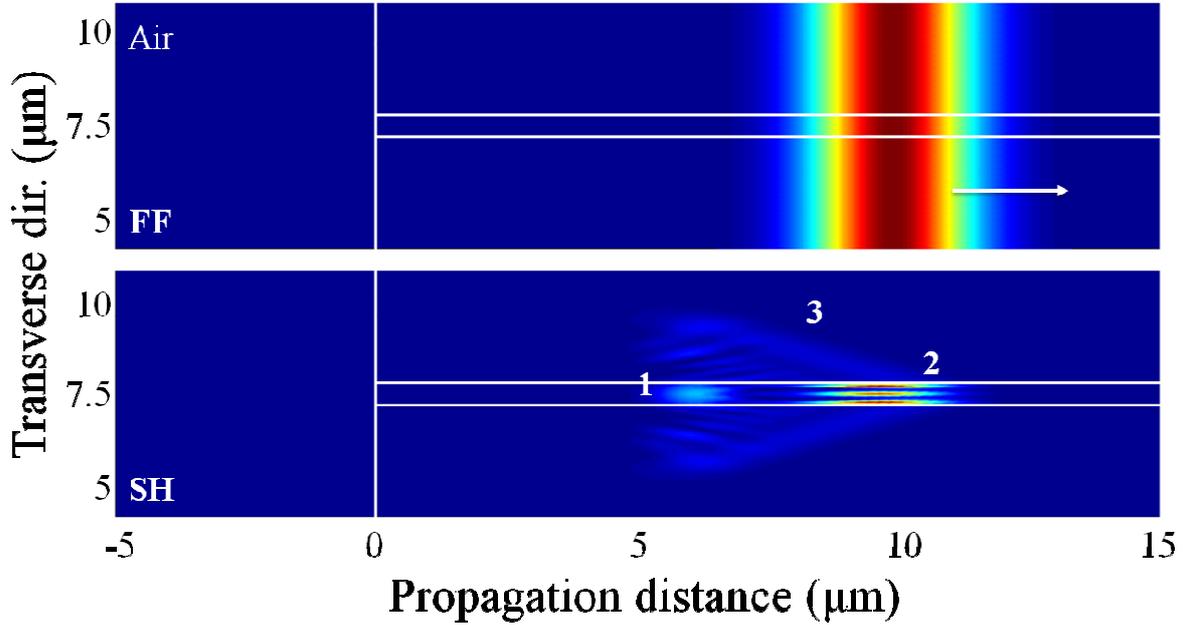

**Figure 2** (Color online) (Media 1). Propagation snapshot of a FF pulse ($|E|^2$, norm.units) impinging on the structure in Fig.1 (a). FF and SH are tuned in the transparent spectral region of the material. The generated SH is composed of: (1) the HOM component that walks off and lags behind the FF pulse; (2) the INH component that locks to the FF pulse; (3) the portion that spills out in the linear region thanks to diffraction and is not bound to the pump.

The overlapping of these two different k-vectors within such a small volume causes the SH at position 2 to have a modulated envelope along the transverse direction (x-axis). The initial energy level of the INH SH component is fixed by the initial conditions, namely the relative values of the refraction indices at the FF and SH wavelengths [4]. Diffraction into the linear region causes some SH energy to be lost. However, it is possible to tune or design working conditions so that the peak of the INH SH field remains nearly constant for almost any desired propagation distance. The conversion efficiency is generally low, of the order of $10^{-9}$-$10^{-11}$ [6, 7], and the FF remains undepleted even after propagation distances much larger than the FF wavelength.

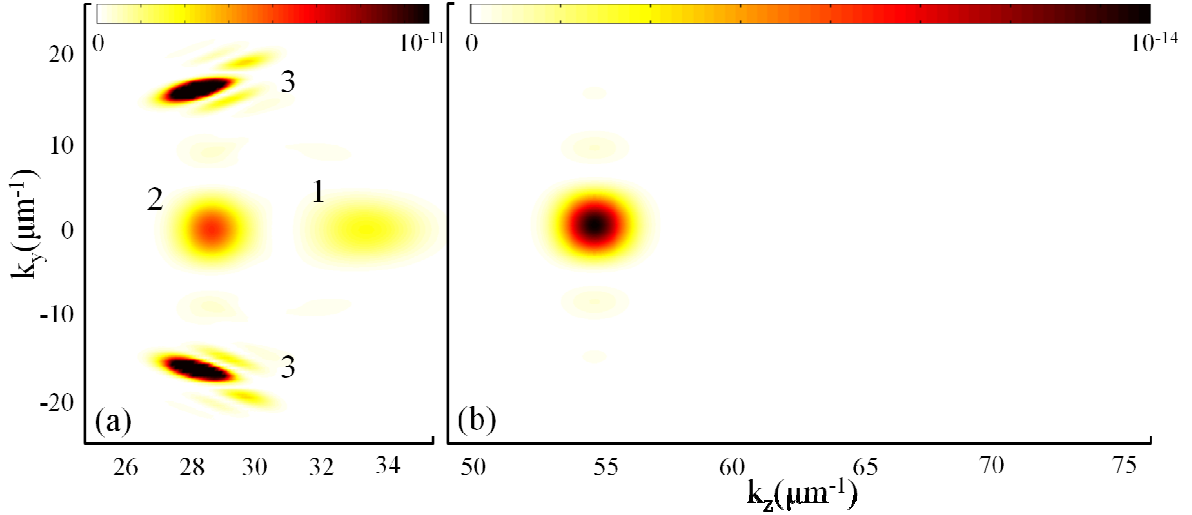

**Figure 3**. (Color online) **(a)** SH Power k-space spectrum (arb.units) of the snapshot in Fig.2. The FF (not shown) has a k-vector in the forward direction (FD) at $k_{FF}=k_\omega n_\omega = 2\pi\omega n_\omega = 14.74\mu m^{-1}$. The SH k-spectrum has three k-components: (1) the HOM component in the FD with $k_{HOM-SH}=2\pi 2\omega n_{2\omega}=32.85\mu m^{-1}$; (2) the INH component in the FD with $k_{INH-SH}= 2k_\omega n_\omega = 4\pi\omega n_\omega=28.9\mu m^{-1}$; (3) the portion that spills out in the linear region with $k_{HOM-SH}= 32.85\mu m^{-1}$. **(b)** SH Power k-space spectrum (arb.units) of the snapshot in Fig.4. FF (not shown) has k-vector in the FD at the value $k_{FF}= k_\omega n_\omega = 27.5\mu m^{-1}$. The SH k-spectrum is now composed only by the INH k-component with $k_{INH-SH}= 2k_\omega n_\omega = 54.9\mu m^{-1}$.

In the next example we tune the FF to $\lambda_{FF}=750$nm so that the material is transparent at the FF, and opaque at the generated SH wavelength (see Fig.1(b)). Fig.4 reports a propagation snapshot in this new environment. Here only the INH SH signal survives, while the HOM SH portion is completely and quickly absorbed. For these tuning conditions $n_\omega=3.28$ and $n_{2\omega}=4.53+i0.44$, resulting in an absorption length of the order of 90nm at the SH wavelength. The envelope of the SH pulse along the z-direction is now a much smoother, Gaussian profile (forged into shape by the FF). The nonlinear waveguide limits the signal along the transverse coordinate. All diffracting portions of the SH signal are quickly absorbed once they spill into the linear region. The k-spectrum in Fig.3(b) clearly shows only one peak in the forward direction at the position $2k_\omega n_\omega$.

Let us now consider a third scenario, where the FF is tuned at $\lambda_{FF}$=500nm and the SH wavelength lies in the spectral region where the material shows a *metallic* behavior (Re($\varepsilon$)<0): although the k-spectrum reveals features almost identical to those shown in Fig.3(b), field propagation phenomena are substantially different. The real part of the permittivity at the SH frequency takes on a negative value, $\varepsilon_{2\omega}$=-5.2+i0.98, while the permeability is µ=1. The sign difference between $\varepsilon$ and µ makes the linear region of the sample forbidden to the HOM SH signal. The linear penetration of the field may be estimated at a mere 9nm. In sharp contrast, the INH SH is trapped by and forced to propagate under the FF pulse, squeezed inside a nonlinear wavelength with pseudo-metallic boundaries.

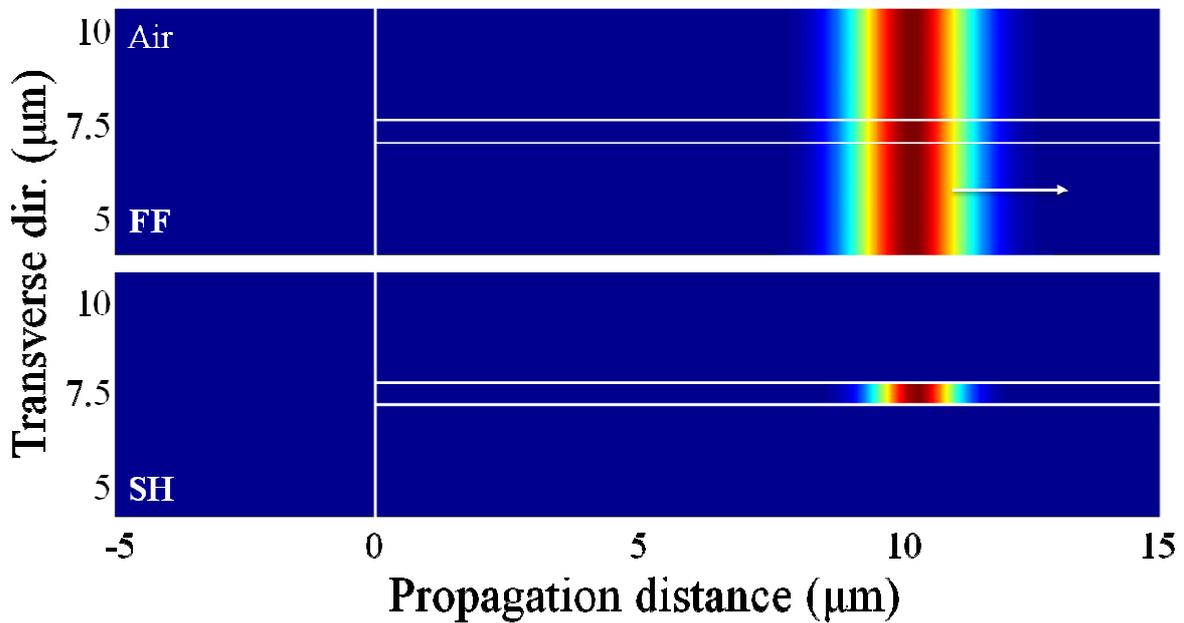

**Figure 4**. (Color online) Same as Fig.1 ($|E|^2$, norm.units) but now the SH is tuned in the absorption spectral region of the medium. All the generated portions of the SH are absorbed but the INH component travels locked to the FF pulse.

The question that one may now ask is: how narrow can one make the waveguide? The main obstacle to any practical use of this device is given by the penetration of the SH field outside the nonlinear region. As we have shown above, this penetration strongly depends on materials and tuning conditions, and can vary from a few nanometers in the *metallic* region, to a few hundred nanometers in the absorptive regime. Even if the practical realization of such a device were to strongly depend on fabrication and material issues, here we present *a key example* of the kind of performance one may come to expect for the geometry we are studying. We designed a sample with two nonlinear waveguides embedded in a medium having the same linear background material and thus the same index of refraction (Fig.5, Media 2). The waveguides are 6nm wide and only 8nm apart. For $\lambda_{FF}$=500nm this represents a geometrical feature of the order of ~$\lambda_{FF}$/62. When the fundamental pulse impinges on the entry interface two SH signals are generated, one in each channel. Since the SH is tuned in a region where the material is *metallic*, only the INH SH component survives. In Fig.5 we monitor the SH momentum density (Poynting vector) given by $S_{SH}$=|ExH|. Negative values of the Poynting vector outside the waveguides indicate strong localizing and confining action on the SH field. No SH is allowed to dwell into the linear region. Thanks to this limitation, the field is perfectly able to resolve the two sub-wavelength spatial features of the waveguide. In this example, the SH propagates 7μm before the peak intensity decays by a factor of 1.5. However the *visibility* [14] of the waveguides is preserved and longer propagation distance can easily be engineered. We stress that this is a polarization independent effect and so similar results hold for TE-polarized fields. Moreover, the confinement effect does not depend on the shape of the waveguide so that any kind of exotic optical circuitry can be envisioned.

In conclusion, we have presented a novel guiding system based on the trapping action of the FF on the generated, INH SH signal. The general validity of this mechanism allows for the design of new waveguiding schemes and devices, with the possibility to circumvent conventional resolution and cut-off limits. This effect can synergistically play its role in concurrence with other linear schemes, for example, in traditional, high-index core linear waveguides, eventually improving their overall flexibility and performances.

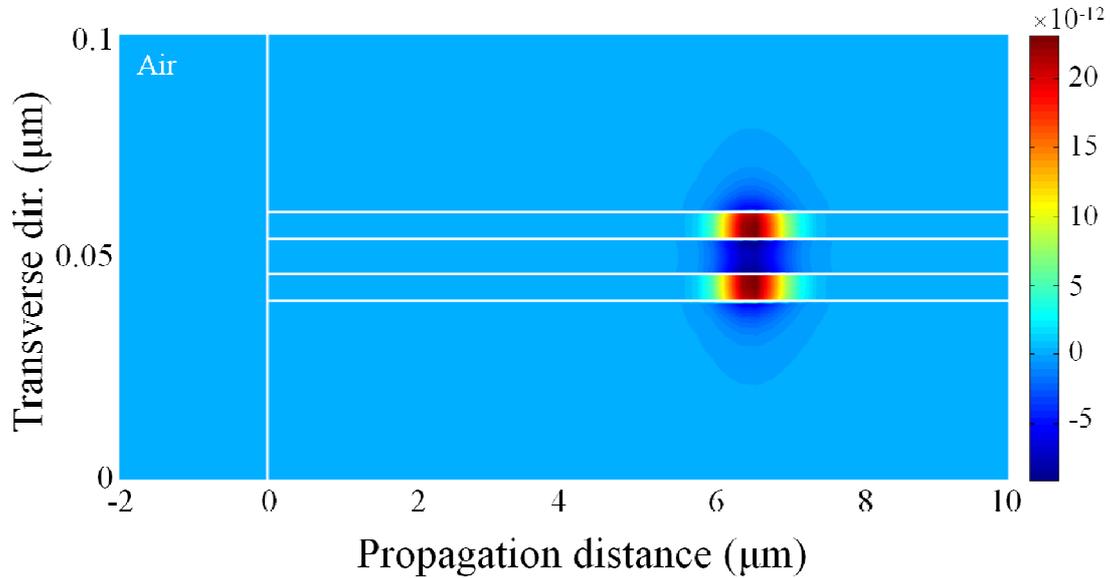

**Figure 5.** (Color online) (Media 2) Propagation snapshot of a generated SH signal (momentum density |E×H|, normalized units) when a FF pulse impinges on a linear material having two ultra-narrow, adjacent waveguides. The negative value outside the waveguide region is a clear indication of strong confinement.